\documentstyle[twocolumn,aps]{revtex}
\tightenlines
\begin{document}

\newcommand{\beq}{\begin{equation}}
\newcommand{\eeq}{\end{equation}}
\newcommand{\bea}{\begin{eqnarray}}
\newcommand{\eea}{\end{eqnarray}}
\newcommand{\ba}{\begin{array}}
\newcommand{\ea}{\end{array}}
\newcommand{\om}{(\omega )}
\newcommand{\bef}{\begin{figure}}
\newcommand{\eef}{\end{figure}}
\newcommand{\no}{\nonumber}
\newcommand{\etal}{{\em et~al }}
\newcommand{\cf}{{\it cf.\/}\ }
\newcommand{\ie}{{\it i.e.\/}\ }
\newcommand{\eg}{{\it e.g.\/}\ }

\title{Reverse reconciliation protocols \\ for
quantum cryptography with continuous variables}
\author{Fr\'ed\'eric Grosshans and Philippe Grangier}
\address{Laboratoire Charles Fabry de l'Institut d'Optique 
(CNRS UMR 8501)
F-91403 Orsay, France}

\maketitle

\begin{abstract}

We introduce new quantum key distribution protocols 
using quantum continuous variables, that 
are secure against individual attacks
for any transmission of the optical line between Alice and Bob. In
particular, it is not required that this transmission is larger than  50 $\%$. 
Though squeezing or entanglement may be helpful, they are not required, and
there is no need for quantum memories or entanglement purification. 
These protocols can thus be implemented using coherent states and homodyne detection, 
and they may be more 
efficient than usual protocols using quantum discrete variables.


\end{abstract}

\section{Introduction}
\label{intro}

\subsection{Coherent QKD protocols}
In the presently very active field of continuous
variable quantum information processing, a stimulating question is whether
quantum continuous variables (QCV) may provide 
a valid alternative to the usual ``single photon"
quantum key distribution (QKD) schemes.
Many recent proposals to use QCV for QKD (for a short review see  \cite{prl})
are based upon the use
of ``non-classical" light beams, such as squeezed light, or 
entangled pairs of light beams.
We have shown recently \cite{prl} that there is actually no need for 
squeezed light : an equivalent level of security may be obtained
by simply generating and transmitting random distributions of coherent states.
More precisely,  we have shown in \cite{prl} that a whole family
of secure protocols can be obtained by using either coherent states,
squeezed states, or entangled EPR beams, provided that the transmission 
of the line is larger than $50 \%$ (\ie the losses are less than 3 dB). 
The security of these protocols is related 
to the no-cloning theorem \cite{cerf,GG01}, and 
non-classical features like squeezing or EPR correlations have no
influence on the achievable secret key rate.
The 3dB loss limit of these cryptography protocols makes 
their security demonstration quite intuitive, but there exist in principle 
multiples ways  for Alice and Bob to go beyond this limit, using \eg
entanglement purification \cite{Dal00}. 

In this paper we present new protocols that are secure 
for any value of the line transmission. The basic idea
is to use {\it reverse} reconciliation protocols, that is, 
Alice will try to guess what was received by Bob,
rather than Bob trying to guess what was sent by Alice. In that
case, Alice can always guess better than the eavesdropper Eve:
this is the basic reason for the security of these new protocols.

\subsection {Direct and reverse reconciliation protocols}

In the first step of a generic QKD  protocol, Alice prepares a quantum state 
and sends it to Bob, who makes a measurement on this state.
Alternatively, Alice and Bob may share two EPR-correlated systems,
and they both make measurement on their parts. 
In order to warrant security, Alice and  Bob 
must randomly choose to use different measurement basis, and the
transmitted data  will be significant only when their basis are compatible.
After the quantum exchange,
they have thus to agree on a common measurement basis, and discard the wrong measurements. 
At the end of this step, Alice and Bob (and the potential eavesdropper Eve) 
know a set of correlated measurements,
that we will call the ``key elements".

As a second step, Alice reveals some randomly chosen  samples of the data that she sent, 
and Bob reveals his corresponding measurements.
These samples allow them to measure some relevant parameters of the quantum channel,
\eg the error rate and the transmission, that is called  ``channel gain" for QCV protocols. 
 Knowing the correlations between their key elements, Alice and Bob 
can evaluate the amount of information they share ($I_{AB}$), and the information
the eavesdropper Eve can have about their values ($I_{AE}$ and $I_{BE}$). Therefore
they can evaluate the size of the secret key they will generate at the end of the protocol.
 If Eve knows too much, the size of this secret key will be 0, 
and Alice and Bob will abort the protocol at this step. 

  Now comes  the crucial step of reconciliation, where
  Alice and Bob will use classical communications to extract a common key 
  from their correlated key elements, revealing as little information as possible to a third party
  ignoring these key elements. This step usually uses parity-based algorithms like Cascade. 
  There are actually two main options for doing the reconciliation \cite{Maurer}:

    {\it Direct Reconciliation (DR).} 
Alice sends correction information and Bob corrects his key elements to have 
    the same values as Alice.
    Alice knows from the previous step the minimum amount of information
    she's got to reveal at this step.
    If the reconciliation protocol is perfect, it keeps $I_{AB}-I_{AE}$ constant. 
    At the end of this step, Alice and Bob know a common bit string of length $I_{AB}$,
    and Eve knows $I_{AE}$ bits of this string (slightly more if the reconciliation protocol is not perfect).
    It will provide a useable secret key if $I_{AB}-I_{AE} > 0$ at the 
    beginning.
 We call  this ``direct reconciliation'' (DR) because Bob is reconstructing
    what was sent by Alice, and the classical information flow at this step has the same direction
    as the initial quantum information flow. 

    Direct reconciliation is quite intuitive, and it was used in the coherent state QCV protocol
that we proposed recently\cite{prl}. However, it is not secure as soon as the quantum channel efficiency
falls below $50 \%$. Intuitively, Eve could simulate the losses by a beam splitter and
    look one output port of this beamsplitter. It seems obvious that, if she keeps the biggest part of the beam
    sent by Alice (\ie if she simulate losses higher than 50\%),
    she can extract more information from this beam than Bob ($I_{AE}>I_{AB}$),
    thus forbidding any secret key generation.
This limitation is actually not specific to QCV :
 a ``direct" version of BB84 would be a protocol where Bob would try to fill the ``empty slots"
    where he did not get any photon. It's straightforward to show that this protocol only works when
    the losses are smaller than 3 dB. 
         
   {\it Reverse Reconciliation (RR).}
   We will thus consider ``reverse" reconciliation (RR)
protocols, where Bob sends correction information and Alice corrects
her key elements to have the same values as Bob.
    Since Bob gives the correction information (also to Eve), this type of reconciliation keeps
    $I_{AB}-I_{BE}$ constant, and will provide a useable key if $I_{AB}-I_{BE} > 0$.
    We call it ``reverse reconciliation'' (RR) because Alice adapts herself to what was
    received by Bob.

    In a noiseless BB84 with finite line transmission, this step corresponds to Bob giving to Alice his ``empty slots" 
    where he did not get any photon, and Alice removing the bits she sent at these slots to have the
    same key. Obviously it is also possible to make a reconciliation protocol 
using two way communications, but it can be shown \cite{opt} that reverse reconciliation is
optimum for a coherent state protocol when there is no excess noise in the transmission line (see below). 
Therefore two-ways protocols will not be considered further in the present paper.

 Finally, the last step of a practical QKD  protocol is that  Alice and Bob perform privacy amplification to
   filter out Eve's information. Since this step is based on an evaluation of  the amount of information collected
by Eve  on the reconciled key, a crucial requirement is to get a bound on $I_{AE}$ for DR, and 
on $I_{BE}$ for RR. For a coherent state protocol, the DR bound was given in ref. \cite{prl},
and leads to a security limit for a $50 \%$ line transmission as said above. We will now establish
the RR bound, and we will show that it is not associated with a minimum value of the line transmission.

\section{Entangling cloner}
\label{ReverseCloning}

\subsection{Definition}

To eavesdrop a reverse reconciliation scheme, as described above, 
Eve needs to guess the results of
Bob's measurement. We will call {\it entangling cloner} a system
allowing her to do so, because this kind of system can be
described a cloner creating two quantum-correlated output: Eve
keeps one of them and sends the other to Bob. Let
$(x_{\text{in}},p_{in})$ be the input quadratures of the entangling
cloner and $(x_B,p_B)$, $(x_E,p_E)$ the quadratures of its two
outputs. A good entangling cloner should minimize the conditionnal
variances \cite{qnd} $V_{x_B|x_E}$ and $V_{p_B|p_E}$.

Alice and Bob should assume Eve uses the best possible entangling
cloner, knowing the Alice-Bob channel quality. This channel can be
described by
\bea
  x_B&=&g_x(x_{\text{in}}+B_x) \\
  p_B&=&g_p(p_{\text{in}}+B_p),
\eea
with ($N_0$ is the shot noise variance)
\bea
  \left<x_{\text{in}}^2\right>=\left<p_{\text{in}}^2\right>&\equiv&VN_0\ge N_0 \\
  \left<B_{x,p}^2\right>&\equiv&\chi_{x,p} \; N_0 \\
  \left<x_{\text{in}}B_{x}\right>=\left<p_{\text{in}}B_{p}\right>&=&0
\eea
\subsection{Heisenberg inequalities on Alice and Eve's conditional variances}
\label{SecHsbgVarCond} For reverse reconciliation protocols, Alice
needs to evaluate $x_B$. Her estimator can be noted $\alpha x_A$,
with $\alpha\in {\bf R}$ .  Eve's estimator
for $p_B$ will be $\varepsilon p_E$. Their error will be 
\bea
  x_{B|A,\alpha}     &\equiv&x_B-\alpha      x_A\\
  p_{B|E,\varepsilon}&\equiv&p_B-\varepsilon p_E.
\eea

The commutator $[x_{B|A,\alpha},p_{B|E,\varepsilon}]$ of these two
quantities is then equal to 
\beq
  [x_B,p_B]
 -\alpha\underbrace{[x_A,p_B]}_0
 -\varepsilon\underbrace{[x_B,p_E]}_0
 +\alpha\varepsilon\underbrace{[x_A,p_E]}_0.
 \eeq
We have therefore
 \bea
 [x_{B|A,\alpha},p_{B|E,\varepsilon}]&=&[x_B,p_B]\\
 \left<x_{B|A,\alpha}^2\right>\left<p_{B|E,\varepsilon}^2\right>&\ge&N_0^2
    \label{ProduitVariances}
 \eea
 The conditional variances obey by definition the following
 relations :
\bea
  V_{x_B|x_A}&=&\min_{\alpha}     \left\{\left<x_{B|A,\alpha}     ^2\right>\right\}\\
  V_{p_B|p_E}&=&\min_{\varepsilon}\left\{\left<p_{B|E,\varepsilon}^2\right>\right\}
\eea
and equation (\ref{ProduitVariances}) leads to
 \beq
\label{HsbgEA}
 V_{x_B|x_A}V_{p_B|p_E}\ge N_0^2
         \text{, i.e. }
         V_{p_B|p_E}\ge\frac{N_0^2}{V_{x_B|x_A}}.
 \eeq
By exchanging the roles of $x$ and $p$ one obtains similarly
 \beq
  V_{p_B|p_A}V_{x_B|x_E} \ge      N_0^2
  \text{, i.e. }
    \label{VBEmintmp}
         V_{x_B|x_E}\ge\frac{N_0^2}{V_{p_B|p_A}}.
 \eeq
These inequalities mean that Alice and Eve cannot jointly
know more about Bob's field than allowed by the Heisenberg principle. 

\subsection{Alice's conditional variance}

If Alice creates the field $(x_{\text{in}},p_{in})$, we can write
 \bea
  x_{\text{in}}&=&x_A+A_x\\
  p_{in}&=&p_A+A_p
 \eea
where $x_A$ ($p_A$)is Alice's best estimation of $x_{\text{in}}$
($p_{in}$) and \beq
  \left<A_x^2\right>=\left<A_p^2\right>\equiv sN_0
 \eeq
where $s$ is the amount of squeezing used by Alice to generate this
field. We have then 
 \beq
  \label{smin}
  s\ge\frac1V.
\eeq
 The correlation coefficients are equal to
\bea
 \left<p_A^2 \right>&=&(V-s)N_0             \\
 \left<p_B^2 \right>&=&G_p(V+\chi_p)N_0       \\
 \left<p_Ap_B\right>&=&g_p\left<p_A^2\right> ,
\eea
and allow us to calculate Alice's conditional variance on Bob's
measurement :
\begin{eqnarray}
V_{p_B|p_A}&=&\left<p_B^2\right>-\frac{\left<p_Ap_B\right>^2}{\left<p_A^2\right>}\nonumber\\
           &=&G_pVN_0+G_p\chi_pN_0\nonumber\\
           & &-G_pVN_0+G_psN_0\nonumber\\
           &=& G_p(\chi_p + s)N_0
            \label{VpBA}
\end{eqnarray}
A similar calculation leads to the symmetric relation
 \beq
 \label{VxBA} V_{x_B|x_A}=G_x (\chi_x + s)N_0
 \eeq
These equations and the constraint (\ref{smin}) on the squeezing
give finally
\bea
  V_{p_B|p_A}&\ge& V_{p_B|p_A,\text{min}}=G_p (\chi_p+\frac{1}V)N_0   \\
  V_{x_B|x_A}&\ge& V_{x_B|x_A,\text{min}}=G_x ( \chi_x+\frac{1}V)N_0
\label{VBAmin}
\eea

\subsection{Eve's conditional variance}
\label{VarianceCondEve}

The output-output correlations of an entangling cloner, described e.g. by
 $V_{p_B|p_E}$, should only depend on the density matrix
 of the field $(x_{\text{in}},p_{\text{in}})$ at its input, and not on the
 way this field was built. The inequality (\ref{HsbgEA}) has thus
 to be fulfilled for every physically allowed value of $V_{x_B|x_A}$,
 given the density matrix of the field $(x_{\text{in}},p_{\text{in}})$.
If we look for a boundary to Eve's knowledge by using eq.(\ref{HsbgEA}), we 
we have thus to use the tightest limit on 
 $V_{x_B|x_A}$, that is given by $V_{x_B|x_A,\text{min}}$ 
according to (\ref{VBAmin}). Obviously the same reasoning holds 
for $V_{p_B|p_A}$, with the corresponding tightest limit $V_{p_B|p_A,\text{min}}$.

We have then
 \beq
 V_{x_B|x_E} \ge        V_{x_B|x_E,\text{min}}=\frac{N_0}{G_p(\chi_p+1/V)}
 \label{VBExmin}
\eeq
 and, similarly
 \beq
  \label{VBEpmin}
 V_{p_B|p_E}\ge      V_{p_B|p_E,\text{min}}=\frac{N_0}{G_x(\chi_x+1/V)}
\eeq

\subsection{Implementation}
\label{VraiCloneurInverse}

In a practical QKD scheme Alice and Bob will give the same roles to
$x$ and $p$. Assuming therefore that $G_x=G_p=G$ and $\chi_x=\chi_p=\chi$,
the two bounds (\ref{VBExmin}, \ref{VBEpmin}) reduce to a single one,
and it is possible to explicitly
 describe an entangling cloner achieving this limit. We will consider here
 only the case  where $G<1$, but the limit is tight for any $G$. The
 entangling cloner can then be sketched as follows :
Eve uses a beamsplitter with a transmission $G$ to split up part of the
Alice-Bob transmitted signal, and she injects into the other input
port a field $E1$, with the right variance
to induce a noise of variance $G\chi N_0$ at Bob's end. One 
has therefore:
\bea
  \left<x_{E1}^2\right> = \frac{G \chi N_0}{1-G} \; \; \; \; \; 
  \left<p_{E1}^2\right> = \frac{G \chi N_0}{1-G}
\eea
 Eve should know the maximum about this injected field $E1$, and
 will therefore use an half-pair of EPR-correlated beams, so that
she does perform an ``entangling" attack. We can then write
 \beq
  x_{E1}=x_{\text{known}}+x_{\text{unknown}}
 \eeq
 where $x_{\text{known}}$ stand for Eve's best estimation of
 $x_{E1}$, given by the measure of its brother-beam, and
 $x_{\text{unknown}}$ stand for the noise she cannot know.
We have
 \bea
  \left<x_{\text{unknown}}^2\right>&=&\frac{N_0^2}{\left<x_{E1}^2\right>}=\frac{(1-G)N_0}{G \chi}\\
  \left<x_{\text{known}}^2\right>&=&\left<x_{E1}^2\right>- \left<x_{\text{unknown}}^2\right>
 \eea

Eve also use an output port of the beamsplitter to measure the
field $E2$, which gives her information about the input field :
 \beq
 x_{E2}=gx_{E1}-\sqrt{1-G}x_{\text{in}}.
 \eeq
 She can cancel a part of the noise induced by $E1$ by substracting
 the part proportional to $x_{\text{known}}$. Thus she knows
  \beq
  x'_{E2}=gx_{\text{unknown}}-\sqrt{1-G}x_{\text{in}}.
  \eeq

 We also have
 \beq
  x_B=gx_{\text{in}}+\sqrt{1-G}x_{E1}.
 \eeq
 where Eve already knows the part proportional to $x_{\text{known}}$,
 injected with $x_{E1}$ and she only needs to guess
  \beq
  x'_B=gx_{\text{in}}+\sqrt{1-G}x_{\text{unknown}}
 \eeq
 from $x'_{E2}$.
 We have therefore
 \beq
  V_{x_B|x_{E1},x_{E2}}=V_{x'_B|x'_{E2}}.
 \eeq
The calculation of the quantities $\left<x_B^{\prime2}   \right>$,
$\left<x_{E2}^{\prime2}\right>$, $\left<x'_{E2}x'_B     \right>$
lead  straightforwardly to the conditional variance :
\beq
 V_{x'_B|x'_{E2}}  = \frac{N_0}{G\chi+G/V} = V_{x_B|x_E,\text{min}}
\eeq
showing that the entangling cloner does reach the lower limit 
of (\ref{VBExmin}, \ref{VBEpmin}).

\section{Reverse cryptography}

  \subsection{Tolerable noise}
 In a reverse quantum cryptography protocol, Eve's power is
 limited by the values of $V_{x_B|x_E,\text{min}}$ and
 $V_{p_B|p_E,\text{min}}$ given by (\ref{VBExmin}, \ref{VBEpmin}). In the
 following, we will assume that a ``perfect" Eve is able to reach that limit:
\bea
   V_{x_B|x_E}&=&V_{x_B|x_E,\text{min}}=\frac{N_0}{G_p(\chi_p+1/V)}\\
   V_{p_B|p_E}&=&V_{p_B|p_E,\text{min}}=\frac{N_0}{G_x(\chi_x+1/V)}
\label{VBE}
\eea

  Reverse cryptography is possible when
\bea
\label{ComparVcnd1}
 V_{x_B|x_E} &>& V_{x_B|x_A}\\
\label{ComparVcnd2} \text{ or } V_{p_B|p_E} &>& V_{p_B|p_A}
\eea
 Combining equations (\ref{VpBA}, \ref{VxBA}) and (\ref{VBE}), 
the above conditions become
 \bea
  (G_x\chi_x+G_xs)(G_p\chi_p+\frac{G_p}V) &<&  1 \\
  (G_p\chi_p+G_ps)(G_x\chi_x+\frac{G_x}V) &<& 1.
\eea
 These inequalities give the general conditions for the
security of a reverse reconciliation protocol.
For simplicity reasons, we will assume in the following  everything
 is symmetric in $(x,p)$, \ie $G_x=G_p=G$ and
 $\chi_x=\chi_p=\chi$, so that these conditions simplify into:
\beq
  (G\chi+G s)(G \chi + G/V) < 1.
\label{CondCryptoGen}
\eeq 
Any experimental implementation of this
 protocol should however estimate these parameters from
 statistical tests, which are likely not to be exactly symmetric.

\subsection{Secret information rates}
\label{SecIRates}
 The conditions (\ref{CondCryptoGen}) can directly be translated into 
an information rate by using Shannon's formula:
 \bea
  I_{BA}&=&\frac12\log_2\frac{\left<x_B^2\right>}{V_{B|A}}\\
  I_{BE}&=&\frac12\log_2\frac{\left<x_B^2\right>}{V_{B|E}}
 \eea
The secret information rate, for a reverse reconciliation protocol
is therefore
 \bea
  \Delta I&=&I_{BA}-I_{BE}=\frac12\log_2\frac{V_{B|E}}{V_{B|A}}\\
  \label{DeltaI}
  \Delta I&=&\frac12\log_2\frac{1}{\left(G\chi+\frac{G}{V}\right)(G\chi+Gs)}
 \eea

 \subsection{High-modulation limit}
 \label{SecHighMod}

 At the high modulation limit, $V\rightarrow\infty$, and for any
 squeezing $s$ the condition (\ref{CondCryptoGen}) becomes 
 \bea
 G\chi(G\chi+Gs)          &<&1\\
 (G\chi)^2+Gs(G\chi)-1&<&0 \\
 \Delta=G^2s^2+4&&
 \eea
 We have therefore
 \beq
  \label{CondCryptos}
   G\chi< \frac{1}2\left(\sqrt{G^2s^2+4}-Gs\right)
 \eeq
 The inequality (\ref{CondCryptos}) gives the maximum tolerable
added noise for a reverse protocol to be secure.
The corresponding limit is less stringent for low $s$
 values, \ie for strong squeezing. The more squeezing Alice uses,
 the more noise reverse cryptography can tolerate.

 Let us consider the case of a lossy transmission line
with $G\le1$. The added noise is always bigger than a
 minimal value, $ G\chi\ge(1-G)$,
 this inequality being saturated if there is no excess noise
(in that case the only added noise is vacuum noise).
 In that case where the noise is only  due to losses, 
 (\ref{CondCryptos}) becomes
 \bea
 1-G&<&\frac{1}2\left(\sqrt{G^2s^2+4}-Gs\right)
 \eea
It is straightforward to check that this inequality
holds for arbitrary high losses ($G\to 0$), even for coherent states
($s=1$). Therefore reverse reconciliation provides a simple way to extend 
the coherent state protocols of ref. \cite{prl} into the high-loss regime. 
\section{Implementations}

In this section we consider protocols with the same quantum communication part
as in ref. \cite{prl}, but we assume that reverse reconciliation is used. 
The main advantage is that the 3dB loss
limit for security goes away as explained above. Here we give more quantitative
estimates of the security thresholds.

\label{SecImpl}

\subsection{EPR vs coherent beams}
If Alice uses EPR beams (or modulates a maximally squeezed beam),
$s=1/V$, but Alice and Bob have information only every second
transmission, since they don't always choose the same measurement
basis.
 \bea
  \Delta I_{\text{EPR}}
   =\frac14\log_2\frac{1}{\left(G\chi+\frac{G}{V}\right)^2} 
  \label{DeltaIEPR}
   =\frac12\log_2\frac{1  }{      G\chi+\frac{G}{V}         }
\eea

If the added noise only comes from losses, $G\le1$ and
 $G\chi=1-G$. In that case equation (\ref{DeltaIEPR}) becomes
\beq
  \Delta I_{\text{EPR,losses}}=\frac12\log_2\frac1{1-G(1-\frac{1}V)}\ge 0
\label{DeltaIEPRv}
\eeq


 For coherent beams, no squeezing is used, therefore $s=1$ and the mutual
 informations are not dependent of the basis choice.
We have thus
\bea
  \Delta I_{\text{coh}}&=&\frac12\log_2\frac1{\left(G\chi+G\frac1{V}\right)(G\chi +G)}\\
  \label{DeltaIcoh}
  \Delta I_{\text{coh}}&=&\Delta I_{\text{EPR}}-\frac12\log_2G(1+\chi)
\eea

Since $G\chi \ge(1-G)$,
  with the equality iff the noise is minimal and $G\le 1$, we obtain
  \beq
  \Delta I_{coh}\le\Delta I_{EPR},
  \eeq
 both secret rates being equal if and only if the noise comes only from
 losses. As in \cite{prl}, squeezing does not improve the secret
 rate for losses only, but this is no more true in presence of excess noise.
\subsection{Strong losses}
\label{FortesPertes}
 Assuming strong losses, $G\ll1$, no excess noise, and a large initial modulation,
eqs. \ref{DeltaIEPRv} and  \ref{DeltaIcoh} become
 \beq
 \Delta I_{\text{EPR,losses}} = \Delta I_{\text{coh,losses}}\simeq \frac{G}{2\ln2}.
 \eeq
 This secret rate can be compared with  BB84's rate, which is
 $\frac12G\bar n$, with $\bar n=1$ for single photons and
 $\bar n\ll1$ for weak coherent pulses. Even if BB84 uses two
 modes of the electromagnetic field, it is slightly less efficient
 than our reversed continuous variable protocols, but the order of
 magnitude is the same (for strong losses).

 Taking for instance a 100~km line with 20~dB
loss ($G=0.01$) and a reasonable modulation ($V\simeq10$), the
secret key rate is $6.5\cdot10^{-3}$ bit/symbol. For the same
parameters, the secret key rate for QDV QKD with an ideal single
photon source would be at best $5\cdot10^{-3}$ bit/time slot, and
would be one order of magnitude smaller using attenuated light
pulses with $\bar n=0.1$, even with perfect detectors. Actually it
is zero with state-of-the-art QDV systems at 1550 nm. It is also
noticeable that with a ``symbol rate''  of a few MHz that should
be easy to achieve, the QCV secret key rate after 100 km is more
than 10 kbits/sec, while it is simply zero for QDV.

More realistically, one should take into account possible excess noise in the line.
Defining the excess noise as $\epsilon = \chi - (1-G)/G$,
it is simple to show that the reverse protocols
are secure as long as as $\epsilon < (V-1)/(2V) \sim 1/2$ for coherent states, and
 $\epsilon < (V-1)/V \sim 1$ for EPR beams. This shows again that it is
possible to use coherent states, though EPR beams are more robust indeed.

\section{Conclusion}

In this paper we have shown that reverse reconciliation protocols
can be used to extract a secret key from the exchange of coherent, squeezed
or EPR beams between Alice and Bob. The key is secure against individual attacks
for any transmission of the optical line between Alice and Bob, provided
that the excess noise (noise beyond the loss-induced vacuum noise)
is not too large. 

Squeezing makes these protocols more robust against
the excess noise, but it is not absolutely required.
It can be shown \cite{opt} that reverse reconciliation is optimum
for a coherent states protocol with no excess noise in the transmission line, but
this is not always the case : for instance,
direct reconciliation may be better for high line
transmission and large excess noise.
It may therefore be possible to optimize further the secret bit rate
by using two-ways reconciliation protocols.

\end{document}